\documentclass[review]{elsarticle}

\usepackage{lineno,hyperref}
\modulolinenumbers[5]

\journal{Journal of \LaTeX\ Templates}

\usepackage{hyperref}
\usepackage[super,sort,compress]{cite}
\usepackage{subfigure}
\usepackage{color}
\usepackage{soul}

\usepackage{amssymb}
\usepackage{booktabs}
\usepackage{graphicx}
\usepackage{bm}
\usepackage{color}
\usepackage{cancel}
\usepackage{slashed}
\usepackage{subfigure}
\usepackage{soul}

\newcommand{\Ignore}[1]{}

\newcommand{\Ket}[1]{\left\vert #1\right\rangle}

\newcommand{\ii}{\mathrm{i}}
\newcommand{\ee}{\mathrm{e}}

\begin{document}

\title{Open multistate Majorana model}

\author{Benedetto Militello}
\address{Universit\`a degli Studi di Palermo, Dipartimento di Fisica e Chimica - Emilio Segr\`e, Via Archirafi 36, 90123 Palermo, Italia}
\address{I.N.F.N. Sezione di Catania, Via Santa Sofia 64, I-95123 Catania, Italia}

\author{Nikolay V. Vitanov}
\address{Department of Physics, Sofia University, James Bourchier 5 Boulevard, 1164 Sofia, Bulgaria}

\begin{abstract}
The Majorana model in the presence of dissipation and dephasing is considered. First, it is proven that increasing the Hilbert space dimension the system becomes more and more fragile to quantum noise, whether dephasing or dissipation are mainly present. 
Second, it is shown that, contrary to its ideal counterpart, the dynamics related to the open Majorana model cannot be considered as the combined dynamics of a set of independent spin-$1/2$ models.
\end{abstract}

\maketitle

\section{Introduction}\label{sec:introduction}

The dynamics of a quantum system governed by a time-dependent Hamiltonian in the presence of an avoided crossing has been separately analyzed by Landau, Zener, Majorana and St\"uckelberg~\cite{ref:Landau,ref:Zener,ref:Majo,ref:Stuck}. They independently found a remarkable formula that evaluates the diabatic transitions due to the occurrence of an avoided crossing, provided the evolution is essentially adiabatic far from the instant of time when the bare (diabatic) energies become degenerate. 
In the Majorana model a spin-$j$ particle is considered rather than the special case of a spin-$1/2$ (or a two-level system) as in the standard Landau-Zener (LZ) model. Nevertheless, like in the standard LZ model, the particle is assumed to be subjected to a magnetic field with a static transverse component and a linearly time-dependent longitudinal one.

The LZ and Majorana model have received a lot of attention over the decades, for several reasons. 
First of all, they are solvable models characterized by a time-dependent Hamiltonian, which is something rare, except for situations where special conditions are satisfied~\cite{ref:Barnes2012,ref:Simeonov2014,ref:Sinitsyn2018,ref:Sriram2005,ref:Owusu2010,ref:Chrusc2015}. 
Second, they apply to several physical scenarios~\cite{ref:Dodo2016,ref:Sun2016,ref:Sin2016,ref:Jingfu2017,ref:NoriPRL2005,ref:NoriSR2014,ref:Song2016}.
Third, in the fully adiabatic limit, they allow for quantum state manipulation based on the adiabatic following of the Hamiltonian eigenstates.
Over the decades, many generalizations of the LZ model have been proposed, theoretically analyzed and experimentally realized~\cite{ref:Fishman1990,ref:Bouwm1995,ref:Vitanov1996,ref:Vitanov1999a,ref:Vitanov1999b,ref:Ishkhanyan2004,ref:Militello2015a,ref:Toro2017,ref:MilitelloPRE2018}.
Multilevel or multistate versions of the LZ problem have also been considered, involving series of binary crossings ~\cite{ref:Demkov1967,ref:Carroll1986a,ref:Carroll1986b,ref:Brundobler1993,ref:Demkov2000}, degenerate levels~\cite{ref:Usuki1997,ref:Vasilev2007} and specific coupling schemes~\cite{ref:Ivanov2008,ref:Shytov2004,ref:Sin2015,ref:Li2017,ref:Sin2017}. 
Experiments have been developed dealing with avoided level crossings in several physical scenarios, even involving more than two states. Some of the relevant physical systems are superconducting artificial atoms~\cite{ref:Berns2008,ref:Xiu2017}, Bose-Einstein condensates~\cite{ref:Dreisow2009,ref:Zenesini2009,ref:Tayebirad2010}, nitrogen-vacancy centers~\cite{ref:Fuchs2011} and semiconductor quantum dots~\cite{ref:Reilly2008}.

Many of the generalizations theoretically analyzed and experimentally realized are variants of the Majorana model.
In order to make more realistic the predictions, classical or quantum noise have been incorporated to such models. Beyond contributions dealing with the effects of quantum noise on physical systems subjected to time-dependent Hamiltonian, especially in the adiabatic limit~\cite{ref:Lidar,ref:Florio,ref:MilitelloPRA2010,ref:ScalaOpts2011,ref:MilitelloPScr2011,ref:Wild2016}, the noisy two-state Landau-Zener model has been analyzed in depth~\cite{ref:Ao1991,ref:Wubs2006,ref:Potro2007,ref:Saito2007,ref:Lacour2007,ref:Nel2009,ref:ScalaPRA2011}. Some noisy three-state versions of the LZ model have been analyzed, too~\cite{ref:Ashhab2016,ref:Militello2019a,ref:Militello2019b}. 
Some important contributions for the noisy Majorana model have appeared over the decades. A dissipative Majorana model involving a non-Hermitian Hamiltonian responsible for a probability loss has been considered~\cite{ref:Ellinas1992}. Kenmoe {\it et al.}~\cite{ref:Kenmoe2013} have analyzed the effects of a parameter fluctuation-induced classical colored noise. The effects of a white Gaussian isotropic (i.e., involving all the spin components with similar weights) noise have been considered by Band and Avishai through a phenomenological master equation~\cite{ref:Band2019}. 
The effects of longitudinal quantum noise (i.e., a spin-environment interaction involving operators commuting with the longitudinal component of the spin, then mainly inducing dephasing) on a spin-$1$ Majorana model have been considered by Saito and Kayanuma~\cite{ref:Saito2002}, while Pokrovsky and Sinitsyn~\cite{ref:Pokrovsky2003} have extended this analysis also to the transverse (i.e., an interaction involving operators which do not commute with the longitudinal component of the spin, then mainly implying dissipation) noise for a spin-$1$ particle and to the longitudinal noise for a spin-$j$ model.  Very recently, the combination of longitudinal and transverse noise for the spin-$1$ Majorana model has been analyzed~\cite{ref:Militello2019c}, showing that the dephasing is less detrimental than dissipation.
Until now, an extensive analysis of both longitudinal and transverse quantum noise is not present for a spin-$j$ with $j>1$. 
Since the Majorana model can be used to describe diverse physical systems, such as for example proper spins, multi-state atoms, artificial atoms, where different types of noise can be relevant, it can be useful to consider both dephasing and dissipation sources.
In this paper we provide such an analysis and compare the implications from both types of noise. Our analysis is based on the Davies and Spohn theory, which is the proper way to deal with quantum noise in the presence of time-dependent Hamiltonians. Moreover, since the very introduction of the Majorana model~\cite{ref:Majo} it has been known that the dynamics for the ideal spin-$j$ model can be decomposed as the dynamics of $2j$ spin-$1/2$ particles, which provides interesting features and technical advantages in the theoretical analysis~\cite{ref:Bloch1945,ref:Cook1979,ref:Hioe1987}. Nevertheless, we show here that when the noisy model is considered such helpful property is lost and, because of the interaction with the environment, the system cannot be reduced to a fictitious set of smaller ones.
Our motivation is then twofold. On the one hand, since experiments with systems with more than two states undergoing level crossings have been developed~\cite{ref:NoriPRL2005,ref:NoriSR2014,ref:Reilly2008}, an extensive analysis of the noise effects is of practical importance, and, specifically, could be of great importance in systems well described by the Majorana model with $j>1$. On the other hand, a theoretical analysis of the factorization problem clarifies that the noisy model has to be considered as a whole and does not allow for the simplification realized through the factorization of the evolutions.

The paper is organized as follows. In the next section we generalize the theoretical treatment exploited for the spin-$1$ case in Ref.~\cite{ref:Militello2019c} and analyze the dependence of the population transfer efficiency on the value of $j$, showing that a higher value of $j$ implies a lower efficiency. In Sec.~\ref{sec:independent} we discuss the reduction of the spin-$j$ model to the $2j$ independent spin-$1/2$ problem. Finally, in Sec.~\ref{sec:conclusions} we discuss the results and give some conclusive remarks.

\section{Dissipative Majorana Model}\label{sec:majomodel}

{\it Ideal model --- } We consider a multistate Majorana model describing a spin-$j$ particle immersed in a magnetic field with a linearly time-dependent $z$-component and a static $x$-component. The relevant Hamiltonian can be written as ($\hbar=1$)
\begin{eqnarray}\label{eq:Hamiltonian_ideal}
\hat{H}(t) = \kappa t \hat{J}_z + \Omega\sqrt{2} \hat{J}_x = \omega(t) \hat{J}_\theta\,,
\end{eqnarray}
with
\begin{eqnarray}
\hat{J}_\theta &=& \cos\theta \hat{J}_z + \sin\theta \hat{J}_x \,, \\
\tan\theta &=& \Omega \sqrt{2}  / (\kappa t) \,, \\
\omega(t) &=& \sqrt{(\kappa t)^2+2\Omega^2}\,.
\end{eqnarray}

This Hamiltonian can be diagonalized through the action of the unitary operator 
\begin{eqnarray}
\hat{U}_y(\theta) = \ee^{\ii \theta \hat{J}_y} \,.
\end{eqnarray}

Since we have:
\begin{eqnarray}
\hat{U}_y(\theta) \hat{H}(t) \hat{U}_y(-\theta) = \omega(t) \hat{J}_z \,,
\end{eqnarray}
the eigenstates of $H$ can be obtained as $\Ket{1,m}_\theta = \hat{U}_y(-\theta) \Ket{1,m}_z$, while the eigenvalues are $m \omega(t)$, with $m=-j, -j+1, ..., j-1, j$.

{\it Quantum noise --- } According to the Davies and Spohn theory~\cite{ref:Davies1978}, the master equation describing the dissipative dynamics of a system subjected to a time-dependent Hamiltonian involves a dissipator which connects the instantaneous eigenspaces of the Hamiltonian. The theoretical basis of such assertion is the assumption of very short bath correlations, since it implies that in a short time window the bath sees the Hamiltonian as essentially frozen, then allowing for a standard derivation of a markovian master equation~\cite{ref:Petru,ref:Gardiner}. Therefore, assuming a flat spectrum for the environment (which guarantees short correlations), and a system-environment interaction Hamiltonian
\begin{eqnarray}\label{eq:SysEnvHam}
H_I = \lambda \, \hat{X} \otimes \hat{B}\,,
\end{eqnarray}
we have the following Markovian master equation:
\begin{eqnarray}
\nonumber
\dot\rho = -\ii [H(t), \rho] 
&+& \sum_{\nu=1}^{2j} \gamma(\nu) [ \,\, \hat{X}(\nu) \rho \hat{X}^\dag(\nu) - \frac{1}{2}\{  \hat{X}^\dag(\nu)\hat{X}(\nu), \rho \} ]  \, \\
&+&  \sum_{\nu=-1}^{-2j} \gamma(\nu) [ \,\, \hat{X}(\nu) \rho \hat{X}^\dag(\nu) - \frac{1}{2}\{  \hat{X}^\dag(\nu)\hat{X}(\nu), \rho \} ]  \,,
\end{eqnarray}
with
\begin{eqnarray}
\gamma(\nu) \propto \left\{ 
\begin{array}{ll}
\lambda^2 |\alpha(\nu)|^2 D(\nu) N(\nu, T) + 1 \qquad & \nu>0 \\
\lambda^2 |\alpha(\nu)|^2 D(|\nu|) N(|\nu|, T)  \qquad  & \nu<0
\end{array}
\right. \,\,\,,
\end{eqnarray} 
where  $N(\nu,T)=(\ee^{-\nu/T}-1)^{-1}$ is the mean number of bath bosons at frequency $\nu$, $D(\nu)$ is the density of bath modes at frequency $\nu$, $\alpha(\nu)$ is the coupling strength of the interaction with a mode of frequency $\nu$ and $T$ is the bath temperature. Moreover, $\hat{X}(\nu) = \sum_{\epsilon'-\epsilon=\nu} \hat{\Pi}_\epsilon \, \hat{X} \, \hat{\Pi}_{\epsilon'}$, with $\hat{\Pi}_\epsilon$ projector to the eigenspace of $\hat{H}$ associated to the energy $\epsilon$, that is $\hat{H}\hat{\Pi}_\epsilon = \epsilon\hat{\Pi}_\epsilon$.
This approach to the study of quantum noise in systems with time-dependent Hamiltonians has been extensively used~\cite{ref:Florio,ref:MilitelloPRA2010,ref:ScalaOpts2011,ref:ScalaPRA2011,ref:MilitelloPScr2011}.
Coherently with the flat spectrum hypothesis, the quantity $\gamma \equiv \lambda^2 |\alpha(\nu)|^2 D(\nu)$ must be independent from $\nu$.

{\it Efficiency of the population transfer --- } A detailed analysis of the effects of quantum noise for the spin-$1$ (ot three-state) Majorana model has been developed in Ref.~\cite{ref:Militello2019c}, where an extensive comparison of the effects induced by dephasing or by dissipation has been carried out. The dissipation has been proven to jeopardize the population transfer more than the dephasing. This behavior is valid also when the dimension of the subspace is higher than three. Here we want to focus right on the dependence of the efficiency of the population transfer with respect to $j$. We assume a process where the system starts in the state $\Ket{j,j}$ and evaluate the final population of $\Ket{j,-j}$, where the system is supposed to be after the adiabatic following that bring population from one of such two states to the other.

\begin{figure}[h]
\begin{center}
\subfigure[]{\includegraphics[width=0.7\textwidth, angle=0]{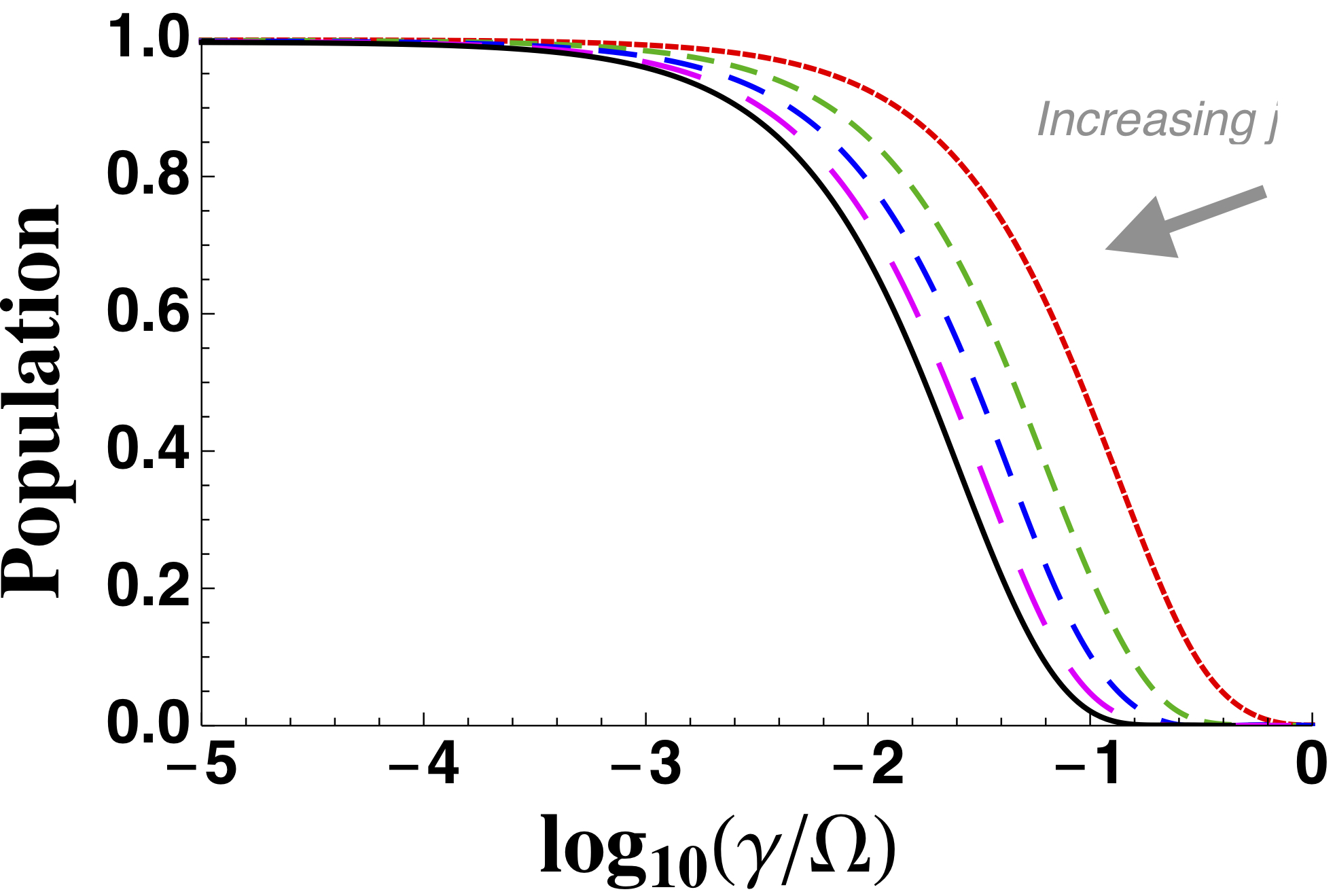}} \\
\subfigure[]{\includegraphics[width=0.7\textwidth, angle=0]{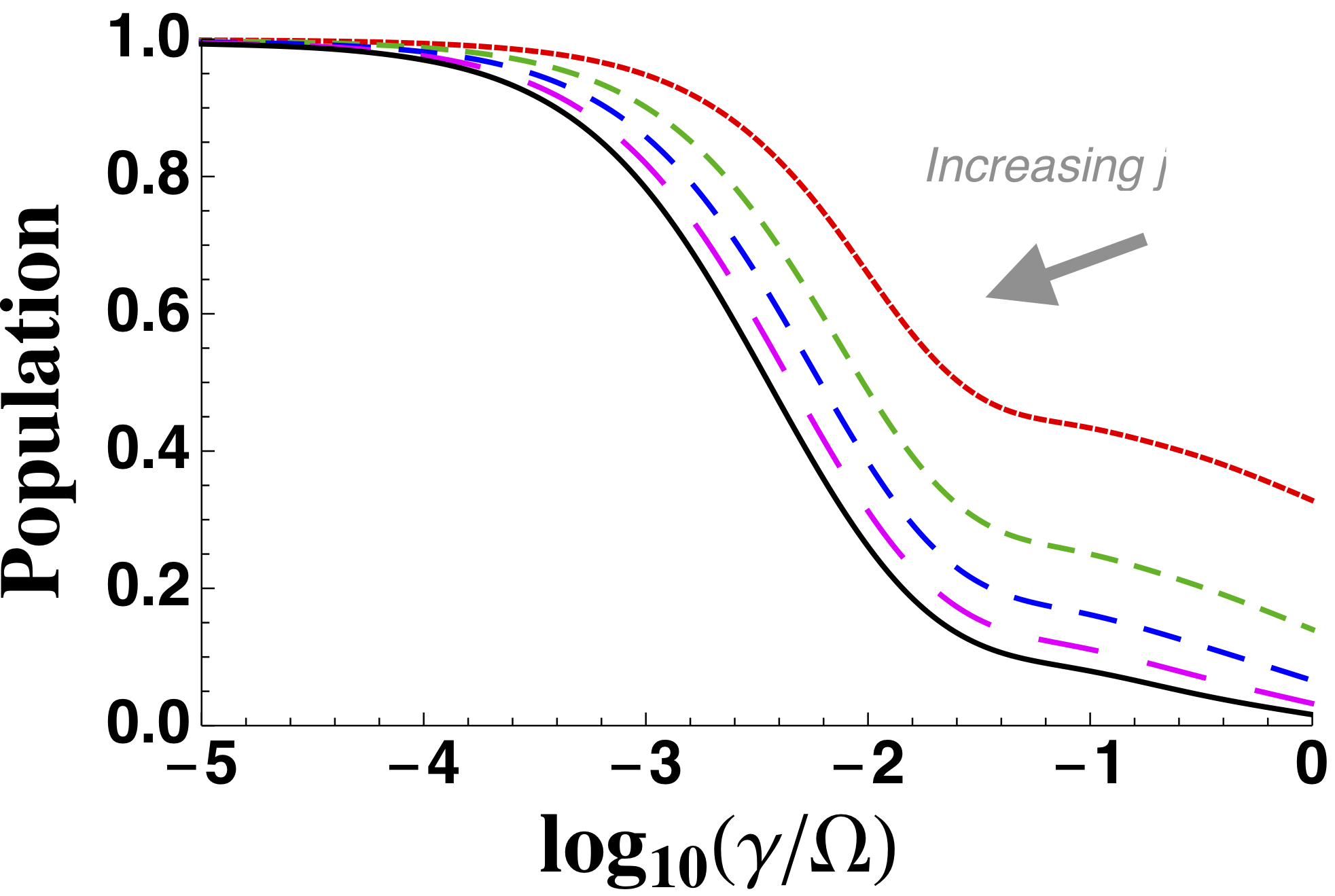}} 
\end{center}
\caption{(Color online) Final population of state $\Ket{j, j}_z$ when the system start in $\Ket{j,-j}_z$ as a function of the system-bath coupling constant $\gamma$ (in units of $\Omega$ and in logarithmic scale). The relevant parameters are: $\hat{X}=\hat{J}_z$, $\kappa t_0 / \Omega = 25$,  $\kappa/\Omega^2=0.1$.  
Two different values of temperature are considered:  $k_\mathrm{B} T / \Omega = 0.001$ (a) and $k_\mathrm{B} T / \Omega = 10$ (b). 
In each plot five different values of $j$ have been considered: $j=1/2$ (dotted red line), $j=1$ (short dashed green line), $j=3/2$ (dashed blue line), $j=2$ (long dashed purple line) and $j=5/2$ (solid black line). It turns out that lower curves correspond to higher values of $j$.} \label{fig:Jz}
\end{figure}

\begin{figure}[h]
\begin{center}
\subfigure[]{\includegraphics[width=0.7\textwidth, angle=0]{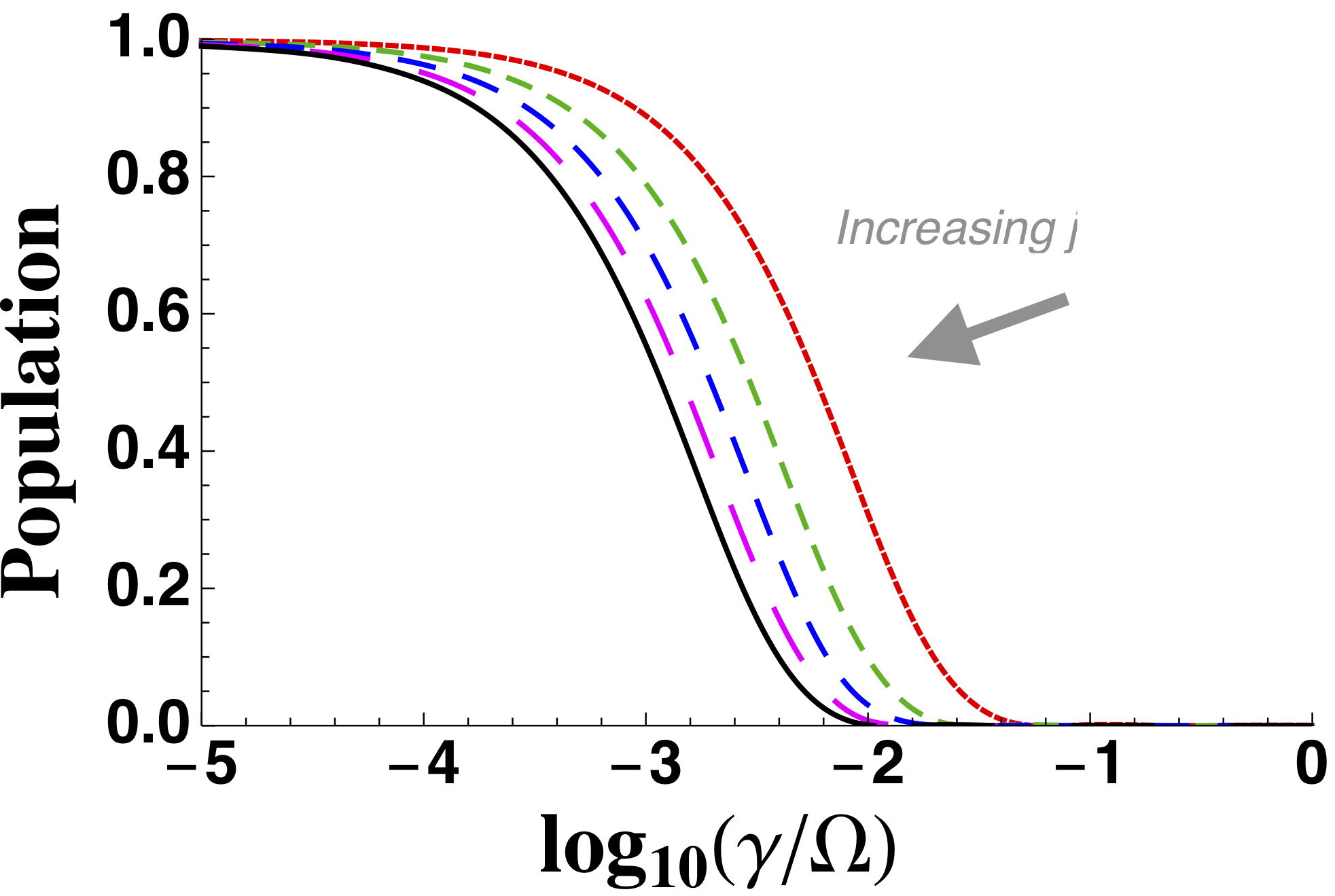}} \\
\subfigure[]{\includegraphics[width=0.7\textwidth, angle=0]{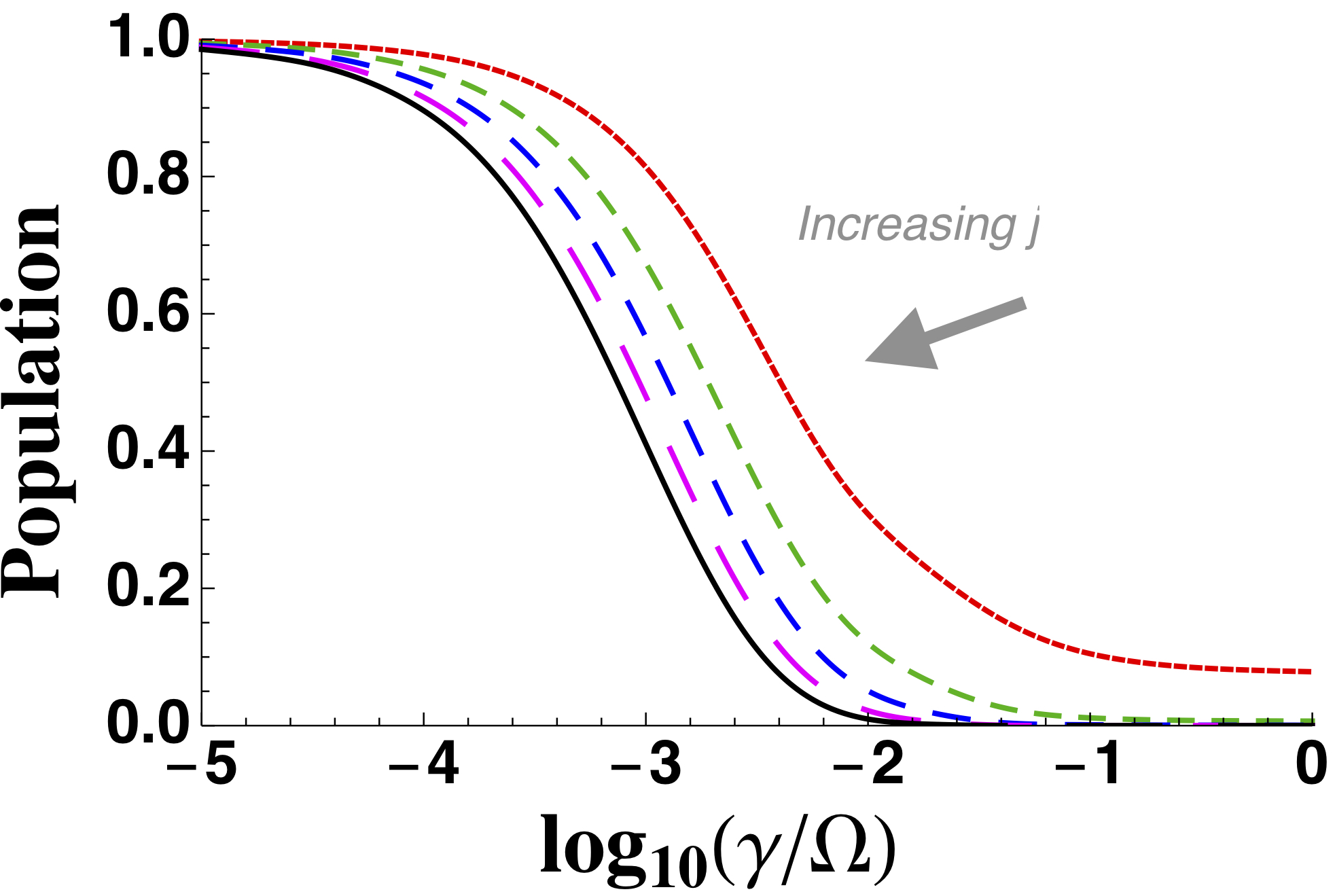}} 
\end{center}
\caption{(Color online) The same as Fig.~\ref{fig:Jz} but for $\hat{X}=\hat{J}_x$.} \label{fig:Jx}
\end{figure}

On the basis of our plots we can assert that for higher values of $j$ the system becomes more sensitive to quantum noise, whether dephasing ($\hat{X}=\hat{J}_z$, in Figs.~\ref{fig:Jz}) or dissipation ($\hat{X}=\hat{J}_x$, in Figs.~\ref{fig:Jx}) is considered, though dissipation clearly has a more detrimental effect. In both figures, we have considered an essentially-zero temperature (\ref{fig:Jz}a and \ref{fig:Jx}a) and a moderately-high temperature (\ref{fig:Jz}b and \ref{fig:Jx}b).

\section{Independent spin-$1/2$ model}\label{sec:independent}

{\it Failure of Majorana's reduction --- } 
A property of the Majorana model for a spin-$j$ particle is that it can be thought of as a model describing the evolution of $N=2j$ independent spin-$1/2$ subjected to the same magnetic field, which has been proven already by Majorana~\cite{ref:Majo} and subsequently studied in depth by several authors~\cite{ref:Bloch1945,ref:Cook1979,ref:Hioe1987}. In fact, given $N=2j$ spin-$1/2$, their sum $\hat{\mathbf{S}}=\sum_{k=1}^{2j} \hat{\mathbf{S}}_k$ is such that $\hat{S}^2$ possesses an eigenvalue $j(j+1)$ in whose subspace the operator $\hat{\mathbf{S}}$ has the same representation of $\hat{\mathbf{J}}$. 
Since the operators associated to different spin-$1/2$ commute, the global evolution operator of the $N$ spins can be cast in the form of a product of single-spin operators:  $\exp(-\ii \mu\, \mathbf{B}\cdot \hat{\mathbf{S}}) = \exp(-\ii \mu\, \mathbf{B}\cdot \sum_{k=1}^{2j} \hat{\mathbf{S}}_k) = \otimes_k \exp(-\ii  \mu\, \mathbf{B}\cdot \hat{\mathbf{S}}_k )$. Moreover, when the system starts in a state which belongs to a subspace of $\hat{S}^2$ the evolution is always confined in such a subspace. This is valid in particular for the subspace corresponding to the eigenvalue $j(j+1)$, and, moreover, in such a subspace $\exp(-\ii \mu\, \mathbf{B}\cdot \hat{\mathbf{S}})$ is equivalent to $\exp(-\ii \mu\, \mathbf{B}\cdot \hat{\mathbf{J}})$. Therefore, after introducing the projector $\hat{\Pi}_j$ onto the subspace with $\hat{S}^2=j(j+1)$, we have $\exp(-\ii \mu\,\mathbf{B}\cdot \hat{\mathbf{J}}) \sim \exp(-\ii \mu\, \mathbf{B}\cdot  \sum_{k=1}^{2j} \hat{\mathbf{S}}_k) \hat{\Pi}_j$.
The independence of the time evolutions can be generalized to the case of a time-dependent magnetic field, by considering the relevant Dyson series. Therefore, the evolution of a spin-$j$ in a magnetic field, whether time-dependent or not, can be thought of as the simultaneous independent evolutions of $N=2j$ spin-$1/2$ in the same magnetic field. 

Once the quantum noise is considered this separation is not possible anymore. In fact, while the unitary part of the dynamics can be separated, the dissipator of a standard master equation in a Lindblad form cannot: $(\hat{A}_1+\hat{A}_2) \rho (\hat{A}_1^\dag + \hat{A}_2^\dag) - \frac{1}{2} \{ (\hat{A}_1^\dag + \hat{A}_2^\dag) (\hat{A}_1 + \hat{A}_2), \rho \} \not=  \sum_k (  \hat{A}_k \rho \hat{A}_k^\dag - \frac{1}{2} \{ \hat{A}_k^\dag \hat{A}_k, \rho \} ) $, no matter if the jump operators $\hat{A}_k$'s commute or not.
Therefore, in general, the evolution associated to an open spin-$j$ Majorana model cannot be recast in the form of $2j$ independent evolutions associated to $2j$ spin-$1/2$ each one interacting with its own bath. 
In other words, while the unitary dynamics can be factorized, the dissipative evolutions cannot.

In many situations a classical noise can be considered instead of a quantum one, because of fluctuations of parameters, such as the amplitude or the phase of an external field. Though a master equation approach is still possible~\cite{ref:Kenmoe2013,ref:Band2019,ref:Saito2002}, sometimes average of solutions of the Schr\"odinger equation is preferred as a technique to evaluate the noise effects~\cite{ref:Laforgue2019}. 
This approach can induce thinking that in the presence of classical noise it is still possible to factorize the dynamics. We show in appendix~\ref{app:ClassicalNoise} that this is not the case, due to some cross correlation terms involved.

{\it Generalization --- } 
The argument for the factorization of the ideal dynamics can be generalized. 
First of all, observe that a given angular momentum operator $\hat{\mathbf{J}}$ for a fixed $j>1$ can be decomposed in several ways as the sum of angular momenta corresponding to smaller values of the the relevant squares. For example a spin-$2$ could be thought of as the sum of four spin-$1/2$, but also as the sum of two spin-$1$ or a the addition of a spin-$1$ and two spin-$1/2$. Each of such schemes gives rise to a different decompositions of the original problem into a set of simpler ones. 
Second, adding every term which is expressible as the sum of terms belonging to the spin Hilbert spaces related to a given decomposition, provided such terms commute with $\hat{J}^2$, will maintain the possibility to decompose the ideal dynamics. However, also in this case it is not guaranteed the possibility to decompose the noisy evolution. To clarify this point, reconsider the specific example of a spin-$2$ and its decomposition as two spin-$1$ problems. Adding terms proportional to the square of the some components of the single angular momenta [say $\propto  (\hat{J}_{1,x}^2 + \hat{J}_{2,x}^2)$, $\hat{\mathbf{J}}_1$ and $\hat{\mathbf{J}}_2$ being the spin-$1$ angular momentum operators], the possibility to decompose the ideal dynamics as the product of two evolutions in spin-$1$ Hilbert spaces will be preserved. Nevertheless, in general, the dissipative evolution will not allow for a similar decomposition.

{\it Failure of an alternative reduction --- } 
There is also another way to think about a multistate Majorana model as an effective two-state model (see Refs.~\cite{ref:Randall2018,ref:Militello2019c}). Indeed, in some cases it is possible to identify some instantaneous Hamiltonian eigenstates which can be expressed as superpositions of two static states, with time-dependent coefficients. For example, for the $j=1$ case, one finds that the middle-energy state can be expressed as $\Ket{1,0}_\theta = \cos\theta\Ket{1,0}_z + \sin\theta (\Ket{1,-1}_z  - \Ket{1,1}_z) / \sqrt{2}$, so that a coherent population transfer from $\Ket{1,0}_z$ to $(\Ket{1,-1}_z  - \Ket{1,1}_z) / \sqrt{2}$, and vice versa, can be realized through the adiabatic following, in the ideal system. 
Nevertheless, as already pointed out in Ref.~\cite{ref:Militello2019c}, the presence of quantum noise implies transitions also to other states, which inevitably prevent the possibility to deduce an effective two-state model.

\section{Conclusions}\label{sec:conclusions}

In this paper we have considered the open Majorana Model, describing a spin-$j$ in a time-dependent magnetic field and in the presence of quantum noise induced by the interaction with an environment. We have assumed an interaction with the environment which involves an angular momentum operator, $\hat{X}$, mainly focusing on the cases $\hat{X}=\hat{J}_z$ and $\hat{X}=\hat{J}_x$. In the former case the system is subjected to dephasing for the majority of the experiment time window. In fact, when the diabatic (bare) energies are very large, the Hamiltonian is almost proportional to $\hat{J}_z$ and then commutes with it. Nevertheless, when the crossing is approached, the role of $\hat{J}_x$ in the Hamiltonian becomes more and more significant to the of becoming dominant, right at the crossing. Therefore, technically speaking, this situation does not involve only dephasing.  Anyway, it is well visible that the system is more sensitive to quantum noise in the mainly dissipative case ($\hat{X}=\hat{J}_x$) than in the mainly dephasing condition ($\hat{X}=\hat{J}_z$). Moreover, we have clearly shown through our numerical resolution of the relevant master equation that for higher values of $j$ the system is more and more sensitive to the effects of the environment. This trend is still valid at nonzero temperature.

Because it has been known since the very Majorana's paper that the spin-$j$ problem can be thought of as $2j$ independent spin-$1/2$ problems, one can be surprised by this $j$-dependence of the noise effects on the system. Nevertheless, by very simple algebraic considerations, it is easy to convince oneself that such a decomposition in spin-$1/2$ dynamical problems is not valid for the open Majorana problem. Indeed, while the spin-$j$ Hamiltonian can be considered as a restriction of the sum of $2j$ spin-$1/2$ Hamiltonians, the dissipator for a spin-$j$ problem cannot be put in the form of $2j$ dissipators of independent spin-$1/2$ systems. In some sense, the environment imposes to see the system as a whole, as it happens in other scenarios where some collective behaviors emerge, such as for example superradiance and subradiance.   

We have also explored the possibility of generalizing the decomposition of the dynamics, using alternative fictitious coupling schemes for the angular momentum and adding appropriate interaction terms. In fact, a given values  of $j>1$ can be obtained in different ways, which allow for different possible factorizations of the dynamics. On this basis, it is easy to show that, adding suitable additional terms in the Hamiltonian, the possibility to decompose the ideal dynamics in terms of simpler ones can be preserved. Nevertheless, we cannot state in general that the dissipative evolution will be decomposable.

\section*{Acknowledgements}

NVV acknowledges support from ERyQSenS (Bulgarian Science Fund Grant No. DO02/3).

\appendix

\section{Failure of Factorization of the Dynamics for Classical Noise}\label{app:ClassicalNoise}

Classical noise due to fluctuations can be incorporated in teh Schr\"odinger equation by adding terms with stochastic coefficients and treating them with the standard perturbation approach followed by an ensemble average. 
Consider a system described by an Hamiltonian obtained as sum of local ones and in the absence of interactions between the susbsystems: $\hat{H}=\sum_k \hat{H}_k + \alpha \eta(t) \sum_k \hat{V}_k $, where $\hat{H}_k$ is the ideal Hamiltonian (in the Majorana model $\hat{H}_k = \kappa t \hat{\sigma}_z + \Omega\sqrt{2}\hat{\sigma}_x$), $\hat{V}_k$ is an interaction terms associated to fluctuating parameters (in the Majorana model it could be proportional to some component of $\hat{\mathbf{S}}_k$), while $\alpha$ is the system-environment coupling strength, and $\eta(t)$ is a stochastic function such that $\langle \eta(t) \rangle = 0$ and $\langle \eta(t) \eta(t') \rangle = \delta(t-t')$. 
The noiseless dynamics described by the unitary operators $\hat{U}_k(t)$ such that $\ii \partial_t \hat{U}_k(t) = \hat{H}_k(t) \hat{U}_k(t)$, while the noisy evolution is described by $\hat{W}_k(t) = \langle \tilde{U}_k (t) \rangle$, where $\ii \partial_t \tilde{U}_k(t) = [\hat{H}_k + \eta(t) \hat{V}_k]  \tilde{U}_k(t)$ and its approximate $n$-th  order (in the parameter $\alpha$) counterpart is denoted by $\hat{W}_{k(n)}$. The whole system operators are $\hat{U}(t)$, $\tilde{U}(t)$, $\hat{W}(t)$ and $\hat{W}_{(n)}(t)$, the first two generated by $\sum_k \hat{H}_k$ and $\hat{H}$, respectively, while the other two are the average of $\tilde{U}(t)$ and result of its $n$-th order truncation.
After passing to the \lq interaction picture\rq\, generated by $\sum_k \hat{H}_k$, we evaluate the effects of quantum noise by using the second order perturbation theory, and then come back to the Schr\"odinger picture:
\begin{eqnarray}
  \nonumber
  \hat{W}_{k(2)}(t) &=& \hat{U}_k(t) \times \left[ 1 - \ii \alpha \int_0^t \mathrm{d}s \, \langle \eta(s)\rangle \, \hat{V}'_k(s) \right. \\
  \nonumber
  &-& \left.  \alpha^2 \int_0^t \mathrm{d}s \int_0^s \mathrm{d}r \, \langle \eta(s) \eta(r) \rangle \, \hat{V}'_k(s) \hat{V}'_k(r) \right] \,,
\end{eqnarray}
which leads to
\begin{eqnarray}
  \nonumber
   \hat{W}_{k(2)}(t) &=& \hat{U}_k(t) \times \left[ 1   {\color{white} \int_0^t } \right. \\   
  \nonumber
  &-& \left. \alpha^2  \int_0^t \mathrm{d}s \int_0^s \mathrm{d}r \, \langle \eta(s) \eta(r) \rangle \, \hat{V}'_k(s) \hat{V}'_k(r) \right] \,,\\
\end{eqnarray}
with $\hat{V}'_k(s) = \hat{U}_k^\dag (s) \hat{V}_k(s) \hat{U}_k (s)$ and where we have suppressed the first order term on the basis of $\langle \eta(s)\rangle = 0$. 
Because of the commutation of operators belonging to different subspaces referring to $k$ and $k'\not=k$, we have: 
\begin{eqnarray}
  \nonumber
  \otimes_k \hat{W}_{k(2)}(t) &\approx& \left[ \otimes_k \hat{U}_k(t) \right] \times \left[ 1 - {\color{white} \int_0^t }  \right. \\
  \nonumber
  &-& \left. \alpha^2 \sum_k \int_0^t \mathrm{d}s \int_0^s \mathrm{d}r \, \langle \eta(s) \eta(r) \rangle \, \hat{V}'_k(s) \hat{V}'_k(r) \right] \,, \\
\end{eqnarray}
where terms of order higher than second have been neglected.

On the contrary, the operator $\hat{W}_{(2)}$ can be straightforwardly evaluated as
\begin{eqnarray}
  \nonumber
  \hat{W}_{(2)}(t) &=& \left[ \otimes_k \hat{U}_k(t) \right] \times \left[ 1 - {\color{white} \int_0^t }  \right. \\
  \nonumber
  &-& \left. \alpha^2 \sum_{k,j} \int_0^t \mathrm{d}s \int_0^s \mathrm{d}r \, \langle \eta(s) \eta(r) \rangle \, \hat{V}'_k(s) \hat{V}'_j(r) \right] \,. \\
\end{eqnarray}
Therefore we get
\begin{eqnarray}
  \nonumber
  \hat{W}_{(2)}(t) &-& \otimes_k \hat{W}_{k(2)}(t) = \otimes_k \hat{U}_k(t) \\
  \nonumber
  &\times& \alpha^2 \sum_{k\not=j} \int_0^t \mathrm{d}s \int_0^s \mathrm{d}r \, \langle \eta(s) \eta(r) \rangle \, \hat{V}'_k(s) \hat{V}'_j(r) \,, \\
\end{eqnarray}
which proves that the product of the independent noisy evolutions and the evolution of the complete noisy system are different.

\end{document}